\documentclass[aps,pra,10pt,superscriptaddress,twocolumn,floatfix]{revtex4-2}
\usepackage{graphicx}
\usepackage{natbib}
\usepackage{amsmath,amsthm,amssymb,dsfont,tabularx,soul}
\usepackage{mdframed}
\usepackage{orcidlink}
\usepackage{hyperref}
\usepackage[switch]{lineno}
\usepackage{xcolor}
\usepackage{soul}
\newcommand{\ket}[1]{\left| #1 \right\rangle}
\newcommand{\bra}[1]{\left\langle #1 \right|}

\newcommand{\braket}[2]{\langle #1|#2 \rangle}
\newcommand{\ketbra}[2]{\left|#1\right\rangle\hskip-1mm\left\langle#2\right|}

\begin{document}
\title{Tracing quantum correlations back to collective interferences}
\author{Ming Ji \orcidlink{0000-0002-6569-5099}}
\email{physmji@gmail.com}
\affiliation{Graduate School of Advanced Science and Engineering, Hiroshima University, Kagamiyama 1-3-1, Higashi Hiroshima 739-8530, Japan}
\author{Jonte R. Hance\,\orcidlink{0000-0001-8587-7618}}
\email{jonte.hance@newcastle.ac.uk}
\affiliation{School of Computing, Newcastle University, 1 Science Square, Newcastle upon Tyne, NE4 5TG, UK}
\affiliation{Quantum Engineering Technology Laboratories, Department of Electrical and Electronic Engineering, University of Bristol, Woodland Road, Bristol, BS8 1US, UK}
\author{Holger F. Hofmann \orcidlink{0000-0001-5649-9718}}
\email{hofmann@hiroshima-u.ac.jp}
\affiliation{Graduate School of Advanced Science and Engineering, Hiroshima University, Kagamiyama 1-3-1, Higashi Hiroshima 739-8530, Japan}
\date{\today}

\begin{abstract}
In this paper, we investigate the possibility of explaining nonclassical correlations between two quantum systems in terms of quantum interferences between collective states of the two systems. We achieve this by mapping the relations between different measurement contexts in the product Hilbert space of a pair of two-level systems onto an analogous sequence of interferences between paths in a single-particle interferometer. The relations between different measurement outcomes are then traced to the distribution of probability currents in the interferometer, where paradoxical relations between the outcomes are identified with currents connecting two states that are orthogonal and should therefore exclude each other.
We show that the relation between probability currents and correlations can be represented by continuous conditional (quasi)probability currents through the interferometer, given by weak values; the violation of the noncontextual assumption is expressed by negative conditional currents in some of the paths. Since negative conditional currents correspond to the assignment of negative conditional probabilities to measurements results in different measurement contexts, the necessity of such negative probability currents represents a failure of noncontextual local realism. Our results help to explain the meaning of nonlocal correlations in quantum mechanics, and support Feynman's claim that interference is the origin of all quantum phenomena.
\end{abstract}
\maketitle

\section{Introduction}

In order to understand the boundary between classical theories and quantum mechanics, we first need to more fully examine the role of quantum correlations. Such correlations are difficult to access directly, because they involve different measurement contexts, defined by mutually incompatible measurements. 
The most well-known example of these correlations is quantum entanglement~\cite{Horodecki2009Quantum}, where the quantum state of one or more degrees of freedom of a system cannot be described independently of the state of one or more degrees of freedom of another system.
These correlations can be observed even when these systems are separated by arbitrarily large distances, and their existence in the quantum formalism is often referred to as nonlocality~\cite{Einstein1935Can}, or nonseparability~\cite{Hance2022Nonlocal,hance2019counterfactualrestrictions}.
The experimental violation of Bell inequalities is the most striking demonstration of the existence of these nonclassical correlations~\cite{Aspect1982Experimental,Rowe2001Experimental,Hensen2015Experimental,Giustina2015Significant,Shalm2015Strong}.
A closely related concept, quantum contextuality, concerns the impossibility of reconciling the statistics observed in different measurement contexts with any context-independent description of the measurement outcomes~\cite{Bell1964On,Kochen1967Problem,Budroni2022ContextualityReview}.
A compact description of the relation between contextuality and nonlocality was first given by Hardy for entangled systems~\cite{Hardy1992Quantum,Hardy1993Nonlocality}, and was later generalised to the relations between measurement outcomes in a three-dimensional space~\cite{Klyachko2008Simple,Cabello2013Simple}.
Cabello and collaborators then proved that contextuality is a generalisation of nonlocality to single-particle systems~\cite{Cabello2010Proposal,Liu2016Nonlocality}.

In order to examine the aspects of the quantum formalism which are behind the paradoxical behaviours associated with nonlocality and contextuality, two of the authors previously analysed the quantitative relations between the different measurement contexts defined by the Hilbert space formalism~\cite{ji2023nonclassical,Ji2024quantitative,hofmann2023sequential}. Ref.~\mbox{\cite{Ji2024quantitative}} showed that quantitative arguments about contextuality based on Hilbert space inner products can be used equally to analyse a three-dimensional Hilbert space of a single system or the nonlocality of two entangled two-level systems. In parallel, Ref.~\mbox{\cite{hofmann2023sequential}} introduced a three-path interferometer, in which the five measurement contexts required for the demonstration of quantum contextuality in a three-dimensional Hilbert space are physically implemented by a sequence of beam splitters. Each of the paths represents a possible measurement outcome obtained in one of the measurement contexts represented by a quantum state, directly relating the problem of contextuality to the problem of attempting to trace the path of a photon through the interferometer. 
Ref.~\mbox{\cite{Ji2024quantitative}} also showed that the nonclassical correlations between two entangled qubits can be expressed in a three-dimensional Hilbert space when collective measurements of the two systems are included in the description. This means that we can map the relations between measurement contexts for qubit pairs onto the path of photon through the three-path interferometer. It is then possible to explore the quantum correlations between two qubits using concepts that were previously limited to the investigation of single-particle contextuality.

Feynman famously claimed that interference is the origin of all quantum phenomena \cite{Feynman2011Vol3}. Ref.~\cite{hofmann2023sequential}, which illustrates contextuality through single particle interferences, provides support for this view. However, it is not immediately obvious how quantum interferences relate to nonclassical correlations between separate quantum systems \cite{Catani2023whyinterference,hance2022comment,catani2022reply,Catani2023Interference}.
Taken together, Ref.~\cite{Ji2024quantitative} (relating contextuality and nonlocality) and Ref.~\cite{hofmann2023sequential} (relating contextuality to interference) suggest a potential route to clarify the relation between nonlocal quantum correlations, and quantum interference between the amplitudes associated with different measurement outcomes. The specific problem that makes it difficult to identify quantum interference effects in both Bell inequalities and Hardy's paradox originates from the assumption of local realism \cite{Budroni2022ContextualityReview}, which requires that the alternative measurement contexts all correspond to local measurements. Focusing on the quantum mechanical description of the relation between two systems, it is important to analyze the meaning of collective measurement outcomes through their expression as a superposition of local states. We can then take a ``shortcut'' through Hilbert space by using the outcomes of collective measurements to relate local measurement outcomes to each other. The physics of nonclassical correlations between two systems can thus be described by the different ways in which the same collective state can be expressed by different superpositions of local states.

This paper is structured as follows.
In Section \ref{sec:II}, we consider a product Hilbert space describing a pair of two-level systems; each measurement outcome can be identified with a state in this product space.
We map the relation between different measurement contexts for this space onto an analogous sequence of interferences between paths in a single particle interferometer. We discuss the relations between the contexts described by the beam splitters in the interferometer, and identify the role of collective measurements and their outcomes. 
 
Using this mapping of collective states to paths, in Section \ref{sec:III} we investigate how the outcomes of collective measurements represent paradoxical relations between the results of local measurements. We show that these paradoxical relations are associated with an interference between quantum state components of a specific maximally entangled state. 

In Section \ref{sec:IV}, we confirm the relation between collective interference and nonclassical correlations by varying the quantum coherence of the input state. The results show that quantum interferences re-direct the probability currents in the interferometer, until it is impossible to trace the origin of the output currents back to the corresponding input port.
 
In Section \ref{sec:V}, we relate the distribution of probability currents to the quasiprobabilities given by the Kirkwood-Dirac distribution~\cite{Kirkwood1933KD,Dirac1945KD,Halpern2018KD}, allowing us to identify the effects of interferences directly with nonclassical correlations between different measurement contexts. Paradoxical statistics are described by negative weak values, corresponding to negative joint probabilities in the Kirkwood-Dirac distribution~\cite{johansen2007quantum,Hofmann2012ComplexJointProbs,Lundeen2012Direct,Bamber2014Dirac,Arvidsson-Shukur2020KD}. We show that the paradoxical correlations associated with the entanglement of the two qubits can be traced back to a negative conditional current at a collective measurement outcome. It is possible to explain the role of these collective measurement outcomes by considering the local states to which they are connected by the beam splitters in the interferometer. Each collective measurement outcome can be identified with a set of conditional statements about the two qubits that must be valid if that outcome is detected. The appearance of negative conditional currents shows that it is impossible to assign simultaneous reality to both the collective statements and the local statements represented by the different measurement contexts. Quantum interferences thus trace the origin of quantum correlations characteristic of entangled states back to the fundamental dependence of the reality of the measurement outcomes on the context established by the application of a specific measurement.
Finally, in Section \ref{sec:VI}, we discuss the implications of our work, and summarise the paper.

\section{Illustration of collective interference using an analogous interferometer}\label{sec:II}

Consider a product Hilbert space describing a pair of two-level systems, where each measurement outcome can be identified with a state in this product space. What is the relation between measurements performed separately on the two systems and collective measurements performed on both systems at the same time? To investigate this question, we start by considering a local measurement, with outcomes $\{\ket{0,0},\ket{0,1},\ket{1,0},\ket{1,1}\}$. These outcomes form a complete orthogonal basis of the Hilbert space of the two systems. Any transformation to a different measurement basis can be represented by interferences between these basis states. We can therefore map the relation between different measurement contexts onto an analogous sequence of interferences between paths in a single particle interferometer. 

First, let us consider the interference between the paths representing the measurement outcomes $\ket{0,0}$ and $\ket{1,0}$ at a beam splitter of reflectivity $R_{01}$. This interference corresponds to a unitary transformation of the measurement basis given by
\begin{equation}
\label{eq:reflectivity}
\begin{split}
\ket{a,0}&=\sqrt{R_{01}}\ket{0,0}-\sqrt{1-R_{01}}\ket{1,0}\\
\ket{b,0}&=\sqrt{1-R_{01}}\ket{0,0}+\sqrt{R_{01}}\ket{1,0}.
\end{split}
\end{equation}
Note that only two of the four paths representing the initial measurement basis interfere here. Specifically, the interference represents a change of basis states that only transforms the states with system two in state $\ket{0}$. 
This corresponds to a quantum controlled unitary operation, where system two is the control system, and system one is the target system. Even though the new context $\{\ket{1,1},\ket{0,1},\ket{a,0},\ket{b,0}\}$ is defined by product states, it describes a collective measurement, where system two must be measured first to determine the measurement basis used on system one (i.e., system 1 should be measured in ``$a$ or $b$'' if system 2 is in state $\ket{0}$, or instead should be measured in ``0 or 1'' if system 2 is in state $\ket{1}$). It is therefore not possible to describe the interference as an effect confined to only one of the two systems. Instead, we should consider it to be a collective interference, describing correlations between the two systems which do not have a classical analogue. 

\begin{figure}
    \centering
\includegraphics[width=\linewidth]{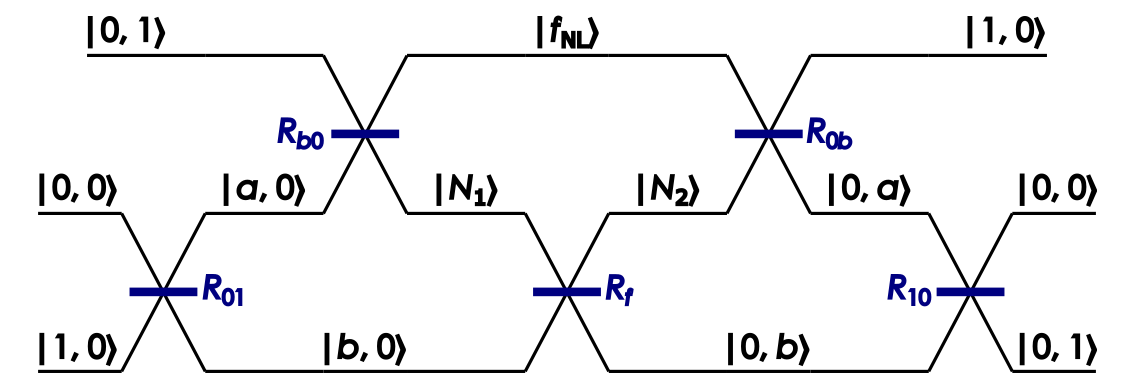}
    \caption{A three-path interferometer, where the paths represent five different measurement contexts being transformed into each other by beam splitters of reflectivities $R_{i}$. The input ports and the output ports represent the same context, but the relative order of the three states has been reversed.}
    \label{fig:ThreePath}
\end{figure}

A local transformation of only one of the two systems would require a corresponding interference between the paths $\ket{0,1}$ and $\ket{1,1}$. However, the collective nature of the interferences between individual pairs of paths allows us to simplify the relations between measurement contexts by excluding the $\ket{1,1}$ path from all the interferences. This means that the $\ket{1,1}$ outcome is shared by all of the measurement contexts. We can then represent the relations between the measurement contexts as a three-path interferometer, where one can always imagine the existence of a fourth path running parallel to the entire interferometer.
To construct a complete set of measurement contexts, sequentially interfere one of the output paths of the immediately preceding interference with the path that was not involved in that interference. This produces a three-path interferometer relating five different measurement contexts to each other, as previously introduced for the investigation of contextuality in a three dimensional Hilbert space ~\cite{hofmann2023sequential}. Fig.~\ref{fig:ThreePath} shows the three-path interferometer applied to the Hilbert space of two separate systems. As explained above, the second beam splitter interferes paths $\ket{0,1}$ and $\ket{a,0}$ with reflectivity $R_{b0}$, where the index $b0$ indicates that the path $\ket{b,0}$ runs parallel to the beam splitter, and is not involved in the interference. Since the output paths cannot be represented by product states, it is not immediately obvious what kind of inter-system relation they describe.

The reflection of $\ket{0,1}$ (transmission of $\ket{a,0}$) forms output path $\ket{f_{\mathrm{NL}}}$, and the reflection of $\ket{a,0}$ (transmission of $\ket{0,1}$) forms output path $\ket{N_{1}}$. 
Thus, both $\ket{f_{\mathrm{NL}}}$ and $\ket{N_{1}}$ can be expressed as a superposition of $\ket{0,1}$ and $\ket{a,0}$.
The precise relation between the two systems described by the corresponding measurement outcomes now depends on the specific reflectivities assigned to the beam splitters. In the following, we consider when $R_{01}=1/2$, so the local measurement bases $\{\ket{0},\ket{1}$ and $\ket{a},\ket{b}\}$ are mutually unbiased. In addition, we require that the state $\ket{f_{\mathrm{NL}}}$ is perfectly symmetric in the two systems, with equal contributions from $\ket{0,1}$ and $\ket{1,0}$. This means the second beam splitter must have a reflectivity of $R_{b0}=1/3$, and identifies the state $\ket{f_{\mathrm{NL}}}$ as an eigenstate of the swap operation $\hat{U}_{\mathrm{SWAP}}$ (which swaps the values of the two qubits), with an eigenvalue of one,
\begin{equation}
\label{eq:symmetryoutcome}
\hat{U}_{\mathrm{SWAP}} \ket{f_{\mathrm{NL}}} = \ket{f_{\mathrm{NL}}}. 
\end{equation}
On the other hand, the output state $\ket{N_{1}}$ is not symmetric in the two systems, and the swap operation changes it into a different state $\ket{N_{2}}$. Using the reflectivities given above, their representations in the input basis read,
\begin{equation}
\begin{split}
&\ket{N_{1}}=\frac{1}{\sqrt{6}}(\ket{0,0}+2\ket{0,1}-\ket{1,0})\\
&\ket{N_{2}}=\frac{1}{\sqrt{6}}(\ket{0,0}-\ket{0,1}+2\ket{1,0}).
\end{split}
\end{equation}
The inner product of these two states is $-1/2$, indicating that the reflectivity $R_{\mathrm{f}}$ of the next beam splitter should be $1/4$, so that the interference between $\ket{N_{1}}$ and $\ket{b,0}$ corresponds to the swap operation acting on the two systems, 
\begin{equation}
    \begin{split}
\ket{N_2}&= \hat{U}_{\mathrm{SWAP}} \ket{N_1}\\
\ket{0,b}&= \hat{U}_{\mathrm{SWAP}} \ket{b,0},
    \end{split}
\end{equation}
The fourth beam splitter with reflectivity $R_{0b}=1/3$ then connects to a context represented by product states, $\{\ket{1,0},\ket{0,a},\ket{0,b}\}$, and the fifth and final beam splitter with reflectivity $R_{10}=1/2$ restores the original context, with paths $\ket{0,1}$ and $\ket{1,0}$ exchanged by the swap operation of the central beam splitter.

$\ket{f_{\mathrm{NL}}}$, $\ket{N_{1}}$ and $\ket{N_{2}}$ cannot be expressed as product states of the two systems. However, the description of these states in terms of interferences relating them to the product states $\{\ket{0,1},\ket{a,0},\ket{b,0}\}$ and $\{\ket{1,0},\ket{0,a},\ket{0,b}\}$ allows us to identify the relations between the two systems described by these collective states. In the next section, we will investigate how these collective measurement outcomes represent paradoxical relations between the results of local measurements.

\section{Relation between local and collective measurement outcomes}\label{sec:III}

As shown in \cite{hofmann2023sequential}, the interferometer given in Fig.~\ref{fig:ThreePath} can be used to demonstrate quantum contextuality, by relating different measurement contexts to each other sequentially. Here, we apply this tool to the problem of nonclassical correlations between two quantum systems. We follow the analogy between contextuality and nonlocality previously established in~\cite{hofmann2023sequential,Ji2024quantitative}. As shown in that work, one could extend the interferometer to fully represent a version of Hardy's paradox. However, we instead intend to show that the three-path interferometer analogy is sufficient to demonstrate collective quantum effects in the two systems, based on the non-separability of the input state. We will do this by investigating how interference describes the relation of the central collective outcome $\ket{f_{\text{NL}}}$ to local outcomes belonging to different contexts.
 
\begin{figure}
    \centering
    \includegraphics[width=\linewidth]{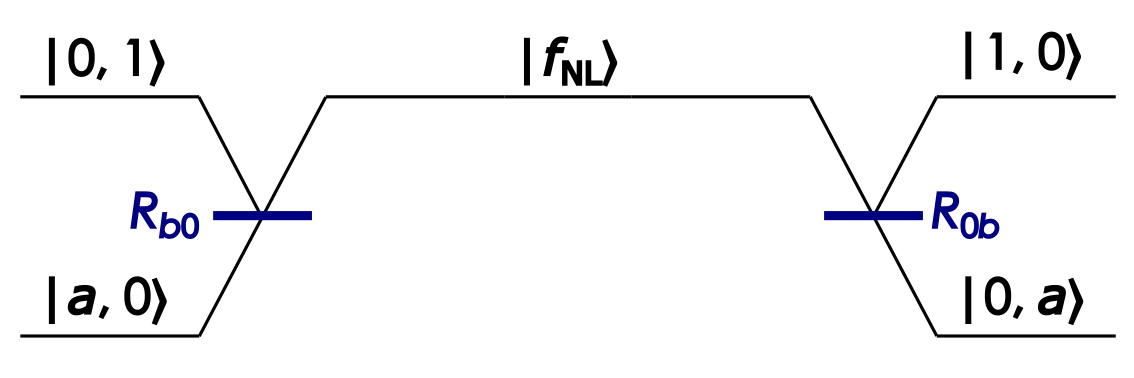}
    \caption{Illustration of the interference effects defining the state $\ket{f_{\text{NL}}}$. The first beamsplitter $R_{b0}$ interferes the paths $\ket{0,1}$ and $\ket{a,0}$, and the second beamsplitter $R_{0b}$ splits $\ket{f_{\text{NL}}}$ into paths $\ket{1,0}$ and $\ket{0,a}.$}
    \label{fig:FNL}
\end{figure}

As shown in Fig.~\ref{fig:ThreePath}, $\ket{f_{\text{NL}}}$ can be related to the interference between the outcomes $\ket{0,1}$ and $\ket{a,0}$ on the one side, and to the interference between the outcomes $\ket{1,0}$ and $\ket{0,a}$ on the other. On the left-hand side, $\ket{f_\mathrm{NL}}$ is described by an interference between $\ket{0,1}$ and $\ket{a,0}$ at a beamsplitter of reflectivity $R_{b0}=1/3$, corresponding to the superposition
\begin{equation}
\label{eq:nonlocalmeasurement1}
\ket{f_{\text{NL}}}=\frac{1}{\sqrt{3}}\ket{0,1}+\sqrt{\frac{2}{3}}\ket{a,0}.
\end{equation}
On the right-hand side, the same collective outcome $\ket{f_{\text{NL}}}$ is split into a coherent superposition of the outcomes $\ket{1,0}$ and $\ket{0,a}$ by a beam splitter of reflectivity $R_{0b}=1/3$, as given by
\begin{equation}
\label{eq:nonlocalmeasurement2}
\ket{f_{\text{NL}}}=\frac{1}{\sqrt{3}}\ket{1,0}+\sqrt{\frac{2}{3}}\ket{0,a}.
\end{equation}
These two relations are illustrated in Fig.~\ref{fig:FNL}.
In noncontextual logic, both superpositions are interpreted as ``either/or'' relations between two logical alternatives. Eqs.~(\ref{eq:nonlocalmeasurement1}) and (\ref{eq:nonlocalmeasurement2}) show that if the outcome $\ket{f_{\text{NL}}}$ is obtained, ``$\ket{0,1}$ or $\ket{a,0}$'' and ``$\ket{1,0}$ or $\ket{0,a}$'' are the only available options in such an interpretation. 
Since $\ket{0,1}$ and $\ket{1,0}$ are mutually exclusive, we can infer two statements from the combination of Eqs.~(\ref{eq:nonlocalmeasurement1}) and (\ref{eq:nonlocalmeasurement2}) and the noncontextual model: 
\begin{enumerate}
\item\label{Statement1} If the noncontextual model predicts the outcome $\ket{f_{\text{NL}}}$, any prediction of $\ket{0,1}$ necessarily implies a prediction of $\ket{0,a}$;
\item\label{Statement2} If the noncontextual model predicts the outcome $\ket{f_{\text{NL}}}$, any prediction of $\ket{a,0}$ necessarily implies a prediction of $\ket{1,0}$.
\end{enumerate}

By combining the two statements, we see that any time the noncontextual model predicts the outcome $\ket{f_{\text{NL}}}$, it also predicts either $\ket{0,a}$, or $\ket{a,0}$, or both. We can express this noncontextual requirement as an inequality of the measurement probabilities for these outcomes,
\begin{equation}
\label{eq:inequality}
P(f_{\text{NL}})\leq P(a,0)+P(0,a).
\end{equation}
Note that this is a typical noncontextual inequality, of the same form as given in~\mbox{\cite{Cabello2008Experimentally,Badziag2009Universality,Kleinman2012Optimal,Pan2013Quantum,Su2015Demonstrating,Kunjwal2015Theorem,Xu2016Reformulating,Krishna2017Deriving,Kunjwal2018From,Schmid2018All,Leifer2020Noncontextuality,ji2023nonclassical,Hance2023Contextuality,Matsuyama2023Quantum}}. This inequality quantifies the limit of the ``either/or" interpretation associated with the detection of single particles. It has been pointed out that its violation can be explained by the intensity distributions of classical wave interferences~\mbox{\cite{Frustagila2016classical}}. As explained in~\mbox{\cite{hofmann2023sequential}}, this is a consequence of the joint presence of all amplitudes and intensities of a classical wave, which makes it impossible to identify the origin of different energy terms in the output. Since a single particle always arrives at only one output port, it is natural to assume that it also entered at only one port. The violation of the inequality indicates that this assumption might be problematic.

It is central that quantum interference can modify the contributions of the different outcomes in a manner that contradicts ``either/or" logic, resulting in a violation of the inequality. Fig.~\ref{fig:FNL} gives an idea of how this violation can be achieved. Quantum interferences should increase the probability of $\ket{f_{\text{NL}}}$ beyond the value of the sum of the probabilities of $\ket{a,0}$ and $\ket{0,a}$. Since the superposition of $\ket{0,1}$ and $\ket{1,0}$ can provide such an enhancement while limiting the probabilities of $\ket{a,0}$ and $\ket{0,a}$ to only half the probabilities of $\ket{0,1}$ and $\ket{1,0}$ respectively, the ideal state for a violation of the inequality should be
\begin{equation}\label{eq:purestatemax}
\ket{\Phi_\mathrm{max}} = \frac{\ket{0,1}+\ket{1,0}}{\sqrt{2}}
\end{equation}
For this state, the probability of $\ket{f_{\text{NL}}}$ is $2/3$, while the probabilities of both $\ket{0,a}$ and $\ket{a,0}$ are both $1/4$.

The state $\ket{\Phi_\mathrm{max}}$ is a maximally-entangled state of the two systems. The state given in Eq.~(\ref{eq:purestatemax}) represents a maximal correlation of outcomes in the $\{0,1\}$ basis. In the collective interference between $\ket{0,1}$ and $\ket{a,0}$, the $\ket{a,0}$ amplitude is simply an attenuated version of the amplitude of $\ket{1,0}$, so the enhanced probability of $\ket{f_{\text{NL}}}$ originates from a constructive interference between $\ket{0,1}$ and $\ket{1,0}$.
Likewise, the amplitude in $\ket{N_1}$ is reduced by destructive interference.
On the other side, a low amplitude in $\ket{N_2}$ interferes with the high amplitude in $\ket{f_{\text{NL}}}$ to concentrate the probability into $\ket{1,0}$, explaining why the probabilities in $\ket{a,0}$ and $\ket{0,a}$ can be lower than allowed by noncontextual models. It is remarkable that the very low amplitudes of $\ket{N_1}$ and $\ket{N_2}$ can each interfere constructively with $\ket{f_{\text{NL}}}$ to restore the high probabilities of $\ket{0,1}$ and $\ket{1,0}$ without having to increase the probability of $\ket{a,0}$ or $\ket{0,a}$. The role of the outcome $\ket{f_{\text{NL}}}$ is to describe a relation between the two systems which, in a noncontextual model, would require a correlation between $\{0,1\}$ and $\{a,b\}$. Quantum interference is necessary to observe independent statistics for $\{0,1\}$ and $\{a,b\}$ when the probability of $\ket{f_{\text{NL}}}$ is larger than $1/2$. 

We can see that there is a quantitative relation between the collective interference between $\ket{0,1}$ and $\ket{1,0}$, and the suppression of correlations between $\{0,1\}$ and $\{a,b\}$, which results in the inequality violation. We will now examine how these quantum interferences modify the correlations associated with the collective measurement outcome $\ket{f_{\text{NL}}}$.  

\section{Collective interference and quantum correlations}
\label{sec:IV}
To fully understand the importance of a collective interference, we need to look at the cases where no such interference happens. For this, we can look first at the cases where the input state is solely either $\ket{0,1}$ or $\ket{1,0}$, before introducing a state with a variable visibility of the interference.

Since the reflectivities $R_{0,1}$ and $R_{1,0}$ are both 1/2,
\begin{equation}
    \begin{split}
        &P(a,0|1,0) = \frac{1}{2},\\
        &P(0,a|0,1) = \frac{1}{2}.
    \end{split}
\end{equation}
For both input state $\ket{0,1}$ and input state $\ket{1,0}$, one of the probabilities on the right-hand side of the inequality in Eq.~(\ref{eq:inequality}) is 1/2, and the other is zero. Therefore, the inequality is satisfied by the probability of $P(f_\mathrm{NL})=1/3$ obtained as a result of the reflectivities of our $R_{0b} = 1/3$ and $R_{b0} = 1/3$. It is worth noting that these states clearly satisfy both Statements 1 and 2 obtained from the noncontextual model---specifically, the state $\ket{0,1}$ corresponds to the prediction in Statement 1, and the continuity in the input-output relations at the beamsplitter $R_{b0}$ requires that the probability of finding $\ket{0,1}$ is equal to the sum of the probabilities of $\ket{f_\mathrm{NL}}$ and $\ket{N_1}$ (and similarly for  $\ket{1,0}$, Statement 2, and $\ket{N_2}$). Note that the continuity of input-output relations at beamsplitters can have similar role to noncontextuality, but may lead to different consequences; this is something we will explore further in Section \ref{sec:V}.

We can now consider a situation where the visibility of the interference between $\ket{0,1}$ and $\ket{1,0}$ is not reduced (e.g. by decoherence effects or dephasing). For this, we consider the general input state $\hat{\rho}(\eta)$, where the parameter $\eta$ describes the visibility,
\begin{equation}
\begin{split}
\hat{\rho}(\eta)=&\frac{\ketbra{0,1}{0,1}}{2} + \frac{\ketbra{1,0}{1,0}}{2} +\\
&\eta\frac{\ketbra{0,1}{1,0}}{2}+\eta\frac{\ketbra{1,0}{0,1}}{2}.
\end{split}
\end{equation}
$\eta$ decreases from 1 to 0 as we go from a pure superposition to an incoherent mixture of the two states. For all of these states, $P(0,a)$ and $P(a,0)$ have the same value, of 1/4 each, indicating that Statements 1 or 2 must be violated whenever the probability of $\ket{f_{\text{NL}}}$ exceeds 1/2. This probability depends on the visibility $\eta$, as given by
\begin{equation}
\begin{split}
\label{eq:keyprobability}
P(f_{\text{NL}})&=\text{Tr}[\hat{\rho}(\eta)(\ket{f_{\text{NL}}}\bra{f_{\text{NL}}})]\\
&=\frac{1}{3}(1+\eta).
\end{split}
\end{equation}
As expected, the value of $P(f_{\text{NL}})$ is $1/3$ whenever the value of $\eta$ is zero, due to the absence of collective interference in the input state. $P(f_{\text{NL}})$ can only be greater than the sum of $P(a,0)$ and $P(0,a)$ when $\eta$ is above 1/2. This shows a threshold amount of collective interference is necessary for the failure of the noncontextual model. Interestingly, even some entangled states can satisfy the noncontextual model.

\begin{figure}
\centering
\includegraphics[width=\linewidth]{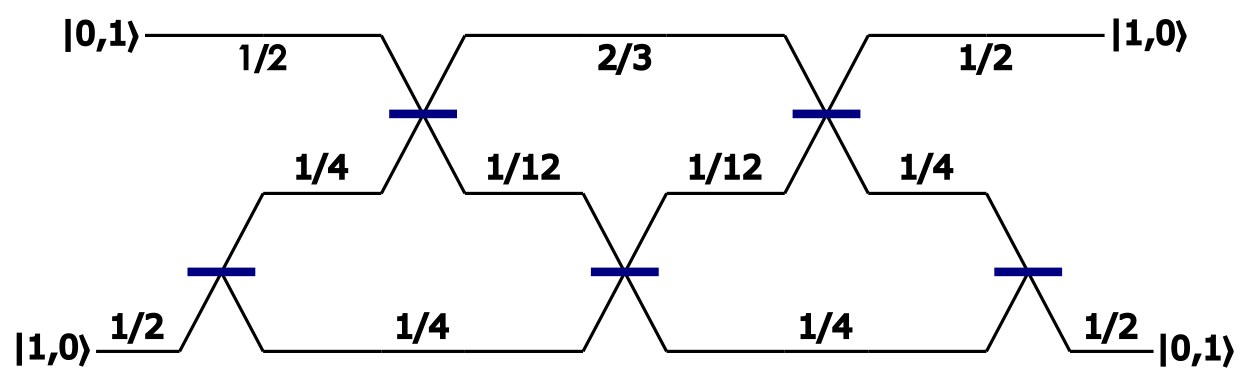}
\caption{Illustration of probabilities in the interferometer for the input state $\ket{\Phi_\mathrm{max}}$. The probability $P(f_{\text{NL}}) = 2/3$ is higher than the sum of probabilities $P(a,0)$ and $P(0,a)$ (1/2). This is equivalent to the observation that the sum of the probabilities of $\ket{a,0}$, $\ket{0,a}$, $\ket{N_1}$ and $\ket{N_2}$ are lower than one.}
\label{fig:Implemented}
\end{figure}

Continuity here requires that, since the $\ket{0,1}$ and $\ket{1,0}$ probabilities are both 1/2, and the $\ket{a,0}$ and $\ket{0,a}$ probabilities are both 1/4, the probabilities of $\ket{N_1}$ and $\ket{N_2}$ are related to the probability of $\ket{f_{\text{NL}}}$ by
\begin{equation}
\label{eq:continuityN1andN2}
    P(N_1) = P(N_2) = \frac{3}{4} - P(f_{\text{NL}}) 
\end{equation}
The interference effect redistributes probability between $P(f_{\text{NL}})$, and $P(N_1)$ and $P(N_2)$. The noncontextual conditions given by Statements 1 and 2 can be satisfied if the current through $\ket{N_1}$ and $\ket{N_2}$ is sufficiently high, to establish a connection between $\ket{b,0}$ and $\ket{1,0}$, and between $\ket{0,1}$ and $\ket{0,b}$, that does not involve either $\ket{a,0}$ or $\ket{0,a}$. If collective interference increases $P(f_\mathrm{NL})$ beyond the threshold of 1/2, it is difficult to identify a continuous probability current through the interferometer representing the collective quantum interference of the two systems. 
The extremal case associated with the input state $\ket{\Phi_\mathrm{max}}$ (where $\eta = 1$) is shown in Fig.~\ref{fig:Implemented}. For this case, the probabilities of $\ket{N_1}$ and $\ket{N_2}$ are each only 1/12. In the figure, we can see it looks as though a current from $\ket{1,0}$ to $\ket{1,0}$ should pass through either $\ket{a,0}$ or $\ket{N_2}$---however, the sums of the probabilities $P(a,0)$ and $P(N_2)$ are not sufficient to account for the probability of obtaining output $\ket{1,0}$. This suggests that part of the probability current in the output port $\ket{1,0}$ should be transferred from the input port $\ket{0,1}$.

Given we expect a continuity of probability, we could ask whether it is possible to trace probability currents through the interferometer, if we relax the noncontextual assumptions in accordance with the effects of quantum interference. Since we have already seen that the correlations between the two systems include collective interference effects that are responsible for the apparent contradiction between these correlations and noncontextual models, we would expect to find an alternative explanation of the nonclassical correlations.

\section{Negative weak values as indicators of contextual continuity}
\label{sec:V}

Our problem in understanding the violation of noncontextual logic originates from an expectation of joint realities of outcomes from different contexts. If we identify these outcomes with the paths taken by a particle in an interferometer, this expectation corresponds to the assignment of a single input and single output path for the particle at each beamsplitter, irrespective of where the particle is eventually detected. However, it is not possible to experimentally observe both an single input and output path for each beamsplitter when interference takes place. Instead, the inequality in Eq.~(\ref{eq:inequality}) refers to the consistency between the individual detection probabilities and an assumed joint probability of finding the particle in a sequence of paths.
However, in quantum mechanics, the assignment of joint probabilities is inherently problematic; an approach taken to address this problem is the introduction of quasiprobabilities that do not conform to the expectation of having real positive values (and so therefore do not necessarily satisfy Kolmogorov's first axiom).
As was shown in~\cite{hofmann2023sequential}, interference effects can be described well using the Kirkwood-Dirac distribution ~\cite{Kirkwood1933KD,Dirac1945KD,Halpern2018KD,Arvidsson-Shukur2020KD}, which is based on weak values \cite{Aharonov1988Weak,aharonov1990properties,Spekkens2008NegCont,Pusey2014Anom,Hance2023Weak} of the projection operators associated with the measurement outcomes. For an input state $\hat{\rho}$, a fixed output $\ket{o}$, the weak value of an outcome $\ket{i}$ is given by
\begin{equation}
\label{eq:weakvalue}
W(i|o)=\frac{\braket{o}{i}\bra{i}\hat{\rho}\ket{o}}{\bra{o}\hat{\rho}\ket{o}},
\end{equation}
In the three-path interferometer, these weak values can be interpreted as a decomposition of the current of probability through $\ket{i}$, conditioned on the output port $\ket{o}$~\cite{hofmann2023sequential}. Like probability currents, these weak values respect continuity: the sum of the weak values entering a beamsplitter will be equal to the sum of the weak values exiting that beamsplitter, just like how the sum of the probabilities entering the beamsplitter will be equal to the sum of the probabilities exiting the beamsplitter.
Weak values also respect the orthogonality relations that express the concept of distinguishability in quantum mechanics (e.g. the weak value of state $\ket{0,1}$, conditioned on the postselection $\ket{1,0}$, will always be zero). This means considering the conditional probability currents given by the weak values provides an advantage over considering only unconditioned probability currents through the interferometer, insofar as they respect our intuitions about both continuity and the noncontextuality of the distinguishability of outcomes. However, different from classical probability currents, weak values are not associated with a joint reality of the outcomes $\ket{o}$ and $\ket{i}$, meaning they do not have to respect noncontextual inequalities. We will therefore use these conditional currents to analyse the violation of noncontextual logic by the collective interference of the two systems.

The inequality violation originates from a redistribution of probability from $\ket{N_{1}}$ and $\ket{N_{2}}$ to $\ket{f_{\mathrm{NL}}}$, as shown in Eq.~(\ref{eq:continuityN1andN2}). In the conditional currents described by weak values, this redistribution of probability can be separated into a redistribution of conditional currents between $\ket{N_2}$ and $\ket{f_{\mathrm{NL}}}$ for $\ket{0,1}$, and between $\ket{N_1}$ and $\ket{f_{\mathrm{NL}}}$ for $\ket{1,0}$. Continuity requires that probability currents which flow from $\ket{f_{\mathrm{NL}}}$ and $\ket{N_2}$ to $\ket{0,1}$ must also pass through $\ket{0,a}$, corresponding to the relation between the weak values given by
\begin{equation}\label{eq:0a01}
    W(f_{\text{NL}}|0,1)+W(N_2|0,1)=W(0,a|0,1)
\end{equation}
Likewise, probability currents flowing from $\ket{1,0}$ to either $\ket{f_{\mathrm{NL}}}$ or $\ket{N_1}$ must pass through $\ket{a,0}$, corresponding to the relation between the weak values given by
\begin{equation}\label{eq:a010}
    W(f_{\text{NL}}|1,0)+W(N_1|1,0)=W(a,0|1,0)
\end{equation}
Eqs.~(\ref{eq:0a01}) and (\ref{eq:a010}) are similar to the conditions in Statements 1 and 2, except the weak values can be negative. If conditional currents were limited to positive values, this would prevent any violation of the inequality in Eq.~(\ref{eq:inequality}). We can now reformulate the inequality in Eq.~(\ref{eq:inequality}) by observing that the probability $P(f_{\mathrm{NL}})$ is given by a weighted sum of the probability currents $W(f_{\text{NL}}|0,1)$ and $W(f_{\text{NL}}|1,0)$, while the probabilities $P(0,a)$ and $P(a,0)$ are completely explained by the corresponding conditional currents $W(0,a|0,1)$ and $W(a,0|1,0)$. The inequality violation can then be traced back to the conditional currents through $\ket{N_1}$ and $\ket{N_2}$ only, 
\begin{equation}
\label{eq:weakinequality}
W(N_{2}|0,1) + W(N_{1}|1,0)\geq0.
\end{equation}
For the state $\hat{\rho}(\eta)$, the two conditional currents in this equation depend on the visibility $\eta$ of the interference between $\ket{0,1}$ and $\ket{1,0}$, as given by
\begin{equation}
\label{eq:currents}
W(N_{2}|0,1)=W(N_{1}|1,0)=\frac{1}{3}\left(\frac{1}{2}-\eta\right).
\end{equation}
This shows that sufficient visibility of interference is necessary for negativity of weak values---specifically, a visibility of at least $\eta=1/2$. This is also the threshold for the violation of the inequality in Eq.~(\ref{eq:inequality}), which is a natural consequence of the fact that the two weak values are equal for this symmetric input state.

\begin{figure}
    \centering
   \includegraphics[width=\linewidth]{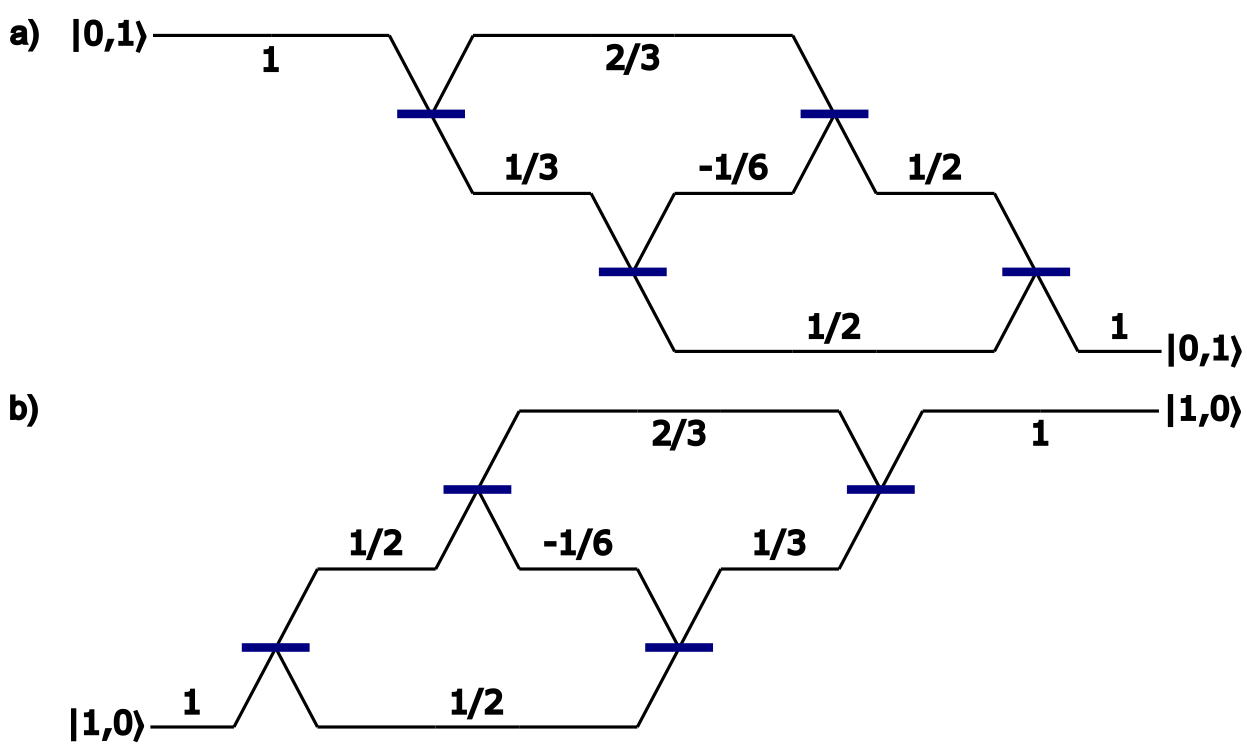}
    \caption{Weak values $W_{\hat{\rho}}(i|o)$ of the projectors on the paths $i$, conditioned by the detected outcomes $o$, for the input state $\ket{\Phi_\mathrm{max}}$. (a) shows the values for $o=\ket{0,1}$, which occurs with a probability of $P(0,1) = 1/2$. (b) shows the values for $o=\ket{1,0}$, which occurs with a probability of $P(1,0) = 1/2$.}
    \label{fig:Only0110Input}
\end{figure}

Fig.~\ref{fig:Only0110Input} shows the  conditional current distributions for the case with maximal visibility of $\eta=1$, giving the negative currents $-1/6$ in paths $\ket{N_1}$ and $\ket{N_2}$. This figure also shows the continuity in the splitting of the path $\ket{0,1}$ into $\ket{f_{\mathrm{NL}}}$ and $\ket{N_1}$ (Fig.~\ref{fig:Only0110Input}a), and merging of the path $\ket{f_{\mathrm{NL}}}$ and $\ket{N_2}$ into $\ket{1,0}$ (Fig.~~\ref{fig:Only0110Input}b). It is interesting to observe that the positive conditional current $W(N_2|0,1)$ overcompensates the negative conditional current $W(N_1|1,0)$, resulting in the directly observable probability of $P(N_1) = 1/12$, and similarly for $W(N_1|1,0)$, $W(N_2|0,1)$, and $P(N_2)$. This ensures that the requirement of positive probabilities is always satisfied when we measure at $N_1$ or $N_2$.

We can now use the relations between outcomes illustrated by the interferometer to explain the correlations between the two systems represented by the complete four-dimensional Hilbert space. For this, we need to understand the physical meaning of the collective measurement outcomes $\ket{f_{\mathrm{NL}}}$, $\ket{N_1}$ and $\ket{N_2}$, in terms of the relation between local properties of the two systems. As we can now see, $\ket{f_{\mathrm{NL}}}$ requires that $\ket{0,1}$ can only occur if $\ket{0,a}$ is true, and $\ket{1,0}$ can only occur if $\ket{a,0}$ is true. This requirement is also valid in the quantum mechanical case, since it does not depend on interferences. $\ket{f_{\mathrm{NL}}}$ makes a joint statement about $\ket{a,0}$ and $\ket{0,a}$, one of which must always be relevant, whether the system is in the state $\ket{0,1}$ or $\ket{1,0}$. The negative conditional currents in $\ket{N_1}$ and $\ket{N_2}$ are needed to offset this relation when the inequality is violated because the probabilities in $\ket{a,0}$ and $\ket{0,a}$ are lower than the probability of $\ket{f_{\mathrm{NL}}}$. In terms of correlations between the two systems, $\ket{N_{1}}$ requires that $\ket{1,0}$ can only occur if $\ket{a,0}$ is true, and $\ket{N_{2}}$ requires that $\ket{0,1}$ can only occur if $\ket{0,a}$ is true. The negative conditional currents of these outcomes represent a suppression of the correlations associated with $\ket{N_{1}}$ and $\ket{N_{2}}$, resulting in a lower value in the probabilities of $\ket{a,0}$ and $\ket{0,a}$. This suppression would not be possible if $\ket{a,0}$ would require that either the correlations described by $\ket{f_{\mathrm{NL}}}$ or $\ket{N_{1}}$ were valid (and same for $\ket{0,a}$, and $\ket{N_{2}}$ or $\ket{f_{\mathrm{NL}}}$). 
If we know something about $\ket{a,0}$, it is not just ignorance that prevents us from applying the conditions expressed by $\ket{f_{\mathrm{NL}}}$ or $\ket{N_1}$. The relation between different contexts cannot be expressed in terms of simultaneous assignment of reality, even if the statements are collective statements about the two systems. Different correlations can seem to contradict each other simply because they refer to incompatible measurement contexts, where collective measurements can represent simultaneous statements about correlations that would be incompatible when measured separately.  We can thus identify the problem of supposed nonlocal interaction (or ``spooky action at a distance''~\cite{Einstein1935Can}) associated with quantum entanglement instead with the fundamental dependence of the reality of the measurement outcomes on the context established by the application of a specific measurement.

\section{Discussion}\label{sec:VI}
The enhancement of probabilities by interference suggests that the idea of a noncontextual reality is invalid in quantum mechanics.
In the case of a single particle, this may not seem to be as extreme as in the bipartite case, as we could always explain the interference process when two paths meet up at a beam splitter in a realist dynamical way. However, the equivalence of the Hilbert space structures of multipartite and single-particle systems indicates that collective interference explains the correlations between separate quantum systems in exactly the same way as single particle interference explains quantum contextuality.
Collective interference explains the absence of a viable local realist model for quantum entanglement by showing that the statistics of the collective system do not correspond to context-independent assignments of reality to measurement outcomes. 
The extreme redistribution of probabilities between measurement outcomes by collective interferences corresponds to negative values in the quasiprobability distributions that describe the correlations between the two entangled systems. 
Collective interference thus describes correlations between different systems in ways which can be illustrated by paths, even though the physics of the interactions between the systems is different from the physics describing the passage of individual quantum particles through beam splitters. In general, the paths in an interferometer represent logical statements which can refer to any physical system, including multipartite systems, despite this difference in their underlying physics. 
This analogue is unique to quantum mechanics, insofar as the unifying Hilbert space structure allows us to map arbitrary statements about complex physical systems to the transmission of particles through a sequence of beamsplitters.
This does not lead to a simple analogy with wave interference, as the separation of wave amplitude and measurement result is nontrivial, even for classical waves \cite{hofmann2023sequential}. The specific quantum-mechanical character of the analogy is expressed by the difference between the meaning of interference for wave amplitudes and for probability amplitudes; the statistical nature of probability amplitudes is necessary for the fundamental contextuality of reality.  

To summarise, we mapped the Hilbert space describing a pair of two-level systems onto an interferometer, where the paths represent a sequence of five different measurement contexts, including collective measurements of the two systems at the centre of the interferometer. 
Using this mapping, we traced the nonclassical correlations described by entanglement between the two systems back to the collective interference between two components of the entangled state. We showed that this quantum interference is responsible for a redistribution of probabilities which cannot be represented by a joint probability distribution for the different measurement contexts. In the interferometer, such a joint probability distribution would describe conditional probability currents travelling from each input port to the corresponding output port. In quantum mechanics, such a continuous expression for conditional currents is provided by the quasiprobability associated with weak values and the Kirkwood-Dirac distribution. While these conditional currents automatically satisfy the expectation of continuity, their values can be negative, explaining the violation of the inequality in Eq. (\ref{eq:inequality}) by entangled states. The magnitude of this violation is directly related to the coherence that defines the entanglement, indicating that the negative conditional currents link the observation of nonclassical correlations directly to collective quantum interferences of the components $\ket{0,1}$ and $\ket{1,0}$.

The results presented here indicate that the nonclassical correlations that characterise entangled states are a natural consequence of collective quantum interferences. These interferences contradict any assignment of noncontextual reality to the outcomes of different measurement contexts, by increasing the probability of the collective measurement outcome $\ket{f_{\text{NL}}}$ beyond the limit of noncontextual models. 
We showed that a visibility of interference above a certain level forces some of the conditional currents to be negative; this visibility threshold is also the threshold at which the system becomes observably contextual. We therefore conclude that the quantum correlations characteristic of entangled states originate from collective interferences that describe a general dependence of reality on the context established by the application of a specific measurement. It is not necessary to invoke any nonlocal interactions in order to explain this phenomenon, since the dependence on the measurement context implies that a joint assignment of reality to the outcomes of different measurements has no physical meaning. Instead, the dependence on measurement context provides an alternative explanation of nonlocal quantum correlations, that supports Feynman's claim that interference is the origin of all quantum phenomena.

\section*{Acknowledgment}
MJ acknowledges support from JST SPRING, Grant Number JPMJSP2132. JRH acknowledges support from Hiroshima University's Phoenix Postdoctoral Fellowship for Research, the University of York's EPSRC DTP grant EP/R513386/1, and the UK's Quantum Communications Hub funded by EPSRC grants EP/M013472/1 and EP/T001011/1.

\bibliographystyle{unsrturl}
\bibliography{ref.bib}

\end{document}